\documentclass[aps,superscriptaddress,twocolumn,twoside,floatfix,pra,nofootinbib,a4paper]{revtex4-2}

\usepackage[charter,cal=cmcal,sfscaled=false]{mathdesign}

\usepackage{dcolumn}
\usepackage{bm}
\usepackage{dsfont}
\usepackage{graphicx,epsfig}
\usepackage{amsmath}
\usepackage{braket}
\usepackage{physics}
\usepackage{caption}
\usepackage{float}
\usepackage[shortlabels]{enumitem}

\usepackage{blindtext}
\usepackage{color}

\usepackage[colorlinks]{hyperref}
\hypersetup{
	colorlinks = true,
	urlcolor = {blue},
	citecolor = {blue},
	linkcolor= {blue}
}

\newenvironment{proof}{%
  \par\noindent\textit{Proof.}%
}{\hfill$\blacksquare$\par}

\renewcommand{\eqref}[1]{Eq.~(\ref{#1})}
\newcommand{\figref}[1]{Fig.~\ref{#1}}
\newcommand{\appref}[1]{App.~\ref{#1}}

\newcommand{\bracket}[3]{\langle#1|#2|#3\rangle}

\newtheorem{result}{Result}

\begin{document}

\title{The role of entanglement in energy-restricted communication
and randomness generation}
\author{Carles Roch i Carceller}\email{carles.roch\_i\_carceller@fysik.lu.se}
\author{Armin Tavakoli}
\address{Physics Department and NanoLund, Lund University, Box 118, 22100 Lund, Sweden.
}

\begin{abstract}
	A promising platform for semi-device-independent quantum information is prepare-and-measure experiments restricted only by a bound on the energy of the communication. Here, we investigate the role of shared entanglement in such scenarios.  For classical communication, we derive a general correlation criterion for nonlocal resources and use it to show that entanglement can fail to be a resource in standard tasks. For quantum communication, we consider the basic primitive for energy-constrained communication, namely the probabilistic transmission of a bit, and show that the advantages of entanglement only can be unlocked by non-unitary encoding schemes that purposefully decohere the entangled state. We also find that these advantages can be increased by using entanglement of higher dimension than qubit.  We leverage these insights to investigate the impact of entanglement for quantum random number generation, which is a standard application of these systems but whose security so far only has been established against classical side information. In the low-energy regime, our attacks on the protocol indicate that the security remains largely intact, thereby paving the way for strengthened security without more complex setups and with negligible  performance reductions.
\end{abstract}

\maketitle

\section{Introduction}
Replacing a classical message with a quantum one can improve the quality of communication between a sender (Alice) and a receiver (Bob). This commonly takes places in a prepare-and-measure (PM) scenario, where Alice encodes her data into a physical system and sends it over a channel to Bob, who decodes it with respect to some property of interest. Since both Alice and Bob are uncharacterised, any quantum advantage must require some form of assumption about the channel. Many specific choices for this assumption have been studied in the literature. For instance, restricting the dimension \cite{gallego2010, Pawlowski2011, tavakoli2018, AlmostQ}, the overlaps of states \cite{brask2017,wang2019}, the fidelity with a target state \cite{tavakoli2021b}, the informational content of the messages \cite{tavakoli2020, Chaturvedi2020}, rotational symmetries \cite{caroline2024} and contextuality \cite{carceller2022}. While some of these approaches are mainly motivated by conceptual interest, their potential for quantum information applications suffers from at least one of two drawbacks: either the assumption itself cannot be directly verified by measurement or there lacks a clear path to high-performance implementation, most relevantly in photonic systems. In response to this, a practically motivated framework has been developed that can overcome these obstacles. It is centered on assuming only  a restriction on the vacuum component of the optical communication \cite{VanHimbeeck2017}. This energy-restricted approach to quantum communication has recently been further developed for entanglement detection \cite{carceller2025depth} and cryptographic tasks \cite{vanhimbeeck2019, Senno2021, Tebyanian2021b, carceller2025photon, carceller2025}, and  a series of recent experiments have realised these ideas, so far with focus on quantum random number generation (QRNG) \cite{Rusca2019, Rusca2020, Tebyanian2021, Avesani2021}.

Common to all these frameworks for semi-device-independent PM scenarios is that they suppose that Alice and Bob do not share entanglement, and sometimes not even classical random variables \cite{Bowles2014, Vicente2017, Tav2020}. Assuming the absence of entanglement is a noteworthy limitation, not only because it is  well-known that entanglement can enhance communication resources  but also because initially unentangled devices can build up entanglement over the many rounds of communication. Only in the dimension-restricted framework has the role of entanglement been explored  \cite{tavakoli2021}, by building on the ideas put forward in the seminal dense coding protocol \cite{bennett1992}. However, it is not given that such ideas will apply to entanglement-assisted prepare-and-measure (EAPM) scenarios based on more physically motivated assumptions than a dimension restriction. Therefore, to pave the way for more practical semi-device-independent quantum information protocols, it is relevant to shed light on EAPM scenarios that operate under energy-restricted communication. As we will show, the role of entanglement is different  and it is no longer guided by the intuition obtained from dense coding.

\begin{figure}
	\centering
	\includegraphics[width=0.85\columnwidth]{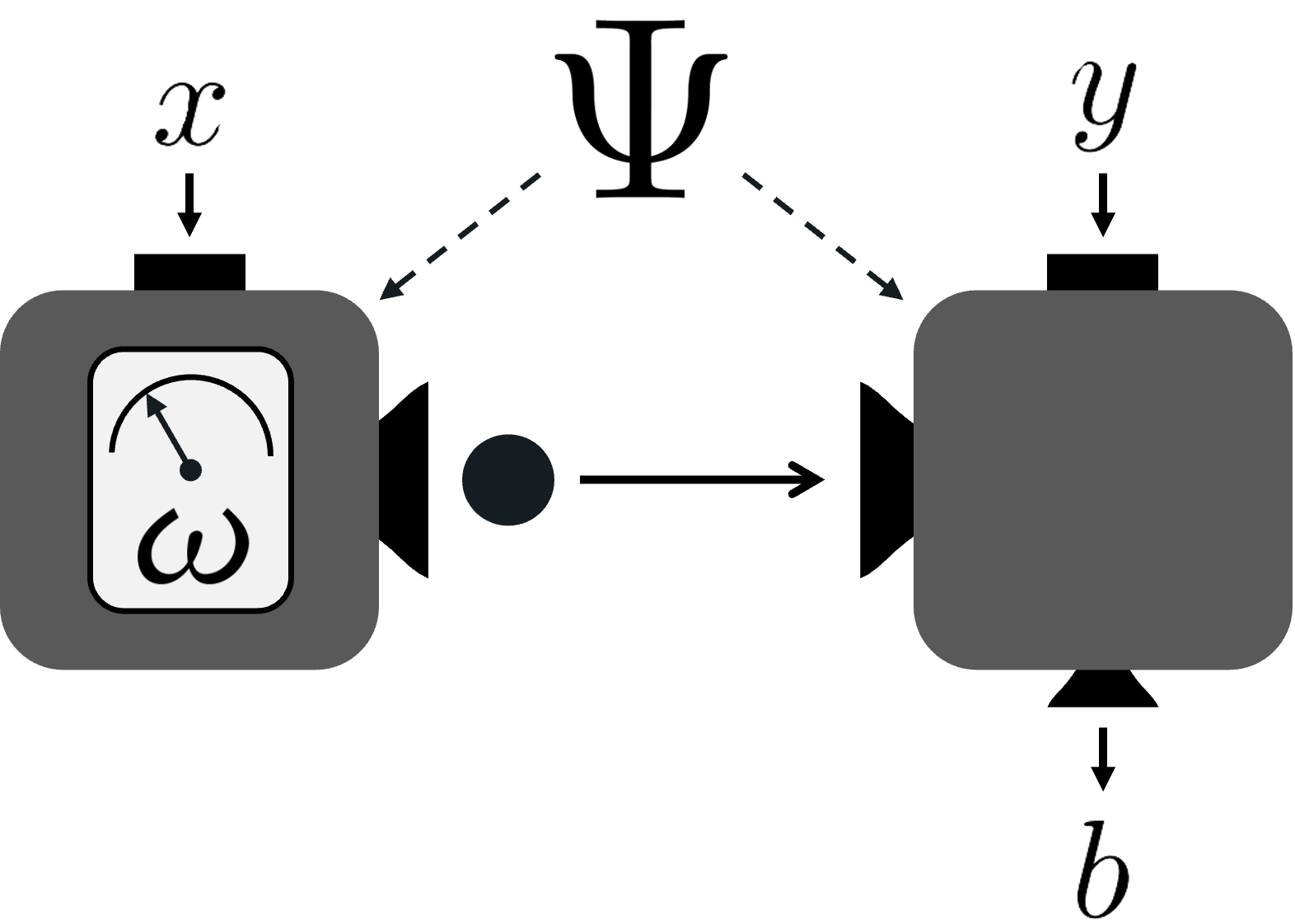}
	\caption{\textit{Entanglement-assisted prepare-and-measure scenario.} Alice and Bob share the state $\Psi$. Alice encodes a classical input $x$ into her state, which is assumed to have no more than $\omega$ units of energy, and sends it  to Bob. Bob uses his input  $y$ to select a measurement whose outcome is outcome $b$. }
	\label{fig:scenario}
\end{figure}

In this work, we investigate EAPM scenarios under energy restrictions when the parties share entanglement, as illustrated in Fig~\ref{fig:scenario}. We first consider classical communication and  derive a simple bound on the correlations achievable when the parties share generic no-signalling resources. Furthermore, we show that classical communication without any shared nonlocal resource saturates this bound in well-known communication tasks, thereby rendering the entanglement resourceless. This contrasts dimension-based scenarios, where entanglement routinely serves as a resource for enhancing classical communication \cite{Buhrman2010}. 

Next, we proceed to the more pertinent scenario in which the communication is carried by a quantum system. We focus on the basic primitive for energy-restricted communication, namely the task of probabilistic transmission of a bit from Alice to Bob, which underpins most proposals in the literature thus far. We prove that no entanglement-assisted protocol based on unitary encodings can provide an advantage over protocols that make no use of entanglement. However, we show that entanglement nevertheless is a resource, but that unlocking it requires non-standard encoding operations that induce noise and thereby decohere the entanglement. While such operations until now have played an at most marginal role in EAPM scenarios, our finding shows that in energy-restricted settings their role is central.

Finally, we deploy these insights for applications in semi-device-independent quantum information. Given the series of experiments focused on energy-restricted QRNG using the probabilistic bit transmission task, we examine possible attacks on such protocols  when the adversary can entangle the devices of the users. This goes beyond the standard security framework for such scenarios, which is based on classical side information \cite{vanhimbeeck2019}. We construct explicit attacks that use entanglement and find that in the high-energy regime this can largely undermine the present security analysis. However, in the low-energy regime, which is most relevant in experiment, we find that the impact of quantum side information is small. This suggests the possibility to make semi-device-independent QRNGs secure against adversaries equipped with entanglement, without increasing the experimental complexity and without noteworthy reductions in the performance.

\section{Scenario and model}

The PM scenario involves two parties, Alice and Bob (see \figref{fig:scenario}). Alice selects classical data, $x$, and encodes it into a message that is transmitted over the communication channel to Bob. Bob selects an input, $y$, with respect to which he decodes the message into an output $b$. When repeated over many identical rounds, the statistics of the experiment are described by the conditional probability distribution $p(b|x,y)$.

A first distinction is between whether the communication channel is classical or quantum. If the communication is classical, the messages are encoded into distinguishable states, e.g.~different symbols or electric pulses. If the communication is quantum, the message is represented by a quantum state that may exhibit superposition features. A second  distinction concerns the pre-established resources between Alice and Bob. One the one hand, Alice and Bob could share a classical random variable, $\lambda$, that allows them to correlate their communication strategy. On the other hand, the most powerful quantum model would permit them to share an entangled state. We now survey the models associated to whether the channel and the shared resource, respectively, are of  classical or quantum nature. We denote the four possibilities as QQ, QC, CQ and CC respectively, where the first letter indicates the nature of the channel and the second the nature of the shared resource.

\subsection{Quantum channel and entanglement}\label{QE}
In the most general setting, which is our main focus here, Alice and Bob share an entangled state $\Psi^{AB}$ and communicate over an energy-restricted quantum channel. This means that Alice uses $x$ to select a quantum encoding operation, that maps her share of $\Psi$ into the quantum message that is sent to Bob. This encoding operation corresponds to a completely positive trace-preserving (CPTP) map, $\Omega^{A\rightarrow C}_x$. The joint state then becomes 
\begin{equation}\label{state}
\tau^{CB}_x=\left(\Omega_x^{A\rightarrow C}\otimes \openone^B\right)[\Psi^{AB}],
\end{equation}
where $C$ denotes the message system. Bob uses $y$ to select a measurement, $\{M_{b|y}\}_b$, where  $b$ denotes the outcome, that is applied to both the quantum message and his share of $\Psi$.  The resulting probability distribution becomes
\begin{align} \label{eq:EAPM_pbxy}
p_{\rm QQ}(b|x,y) = \Tr\left(\tau^{CB}_x M^{CB}_{b|y}\right).
\end{align}
The only restriction imposed in the scenario concerns the vacuum-component of Alice's quantum message, as originally proposed in Ref~\cite{VanHimbeeck2017} for PM scenarios without entanglement. In the EAPM scenario, this means that 
\begin{equation}\label{energycon}
\bracket{0}{\tau^C_x}{0}\ \geq\ 1-\omega, \qquad \forall x ,
\end{equation}
where $\ket{0}_C$ denotes the vacuum state and where $\omega$ is the energy.

\subsection{Quantum channel and shared randomness}
By removing entanglement from the above picture and replacing it with a classical random variable, we reduce the EAPM scenario to the standard energy-restricted PM scenario. This means that $\Psi$  can be viewed as a source of shared randomness, corresponding to selecting $\Psi^{AB}=\sum_\lambda q(\lambda) \ketbra{\lambda}{\lambda}\otimes \ketbra{\lambda}{\lambda}$,  where $q(\lambda)$ is the probability distribution of the classical variable $\lambda$. With this restriction \eqref{eq:EAPM_pbxy} takes the form  
\begin{align} \label{eq:SRPM_pbxy}
p_{\rm QC}(b|x,y) = \sum_\lambda q(\lambda) \Tr\left(\rho_x^{(\lambda)}M_{b|y}^{(\lambda)}\right),
\end{align}
where $\rho_x^{(\lambda)}$ is Alice's quantum message and $M_{b|y}^{(\lambda)}$ is Bob's measurement, both conditioned on the shared variable $\lambda$. The energy constraint in \eqref{energycon} then simplifies to
\begin{equation}
\sum_\lambda q(\lambda)\bracket{0}{\rho_x^{(\lambda)}}{0}\ \geq \ 1-\omega, \qquad \forall x.
\end{equation}

\subsection{Classical channel}
Consider now that the communication channel is classical, but that Alice and Bob still can share entanglement. This means that Alice's encoding operation in section \ref{QE} must output a classical symbol. In other words, the CPTP map $\Omega_x$ becomes a measurement channel,  namely $\Omega_x(\varphi)=\sum_a \Tr\left(\varphi A_{a|x}\right)\ketbra{a}{a}$, for some quantum measurement $\{A_{a|x}\}$ with outcome $a$. The number of possible outcomes is unrestricted. This simplifies the correlations in \eqref{eq:EAPM_pbxy} to
\begin{equation}\label{EACC}
p_{\rm CQ}(b|x,y) = \sum_{a} \Tr\left(A_{a|x}\otimes M_{b|y,a} \ \Psi^{AB}\right) .
\end{equation}
Here, Bob first reads the message of Alice and then uses that knowledge when selecting his measurement \cite{Pauwels2022b}. The energy constraint  reduces to a bound on the probability that Alice produces a null symbol,
\begin{equation}\label{Cecon}
p_A(a = 0|x)\ \geq\ 1- \omega, \qquad \forall x,
\end{equation}
where $p_A(a|x)=\Tr\left(A_{a|x}\otimes \mathds{1}_B \ \Psi^{AB}\right)$ is Alice's message  distribution.

Finally, we consider the fully classical model, where both the channel and the shared resource is classical. This corresponds to Alice and Bob sharing the classical variable $\lambda$ subject to the probability distribution $q(\lambda)$. When Alice reads $\lambda$ she prepares the state  $\rho_x^{(\lambda)}$. Since these states are classical, they are all diagonal in the same basis, namely $[\rho_x^{(\lambda)},\rho_{x'}^{(\lambda')}]=0$ $\forall x,x',\lambda,\lambda'$. The correlations become
\begin{align}\label{fullyC}
p_{\rm CC}(b|x,y) = \sum_{a}\sum_{\lambda} q(\lambda) p_A^{\lambda}(a|x) p_B^{\lambda}(b|y,a),
\end{align}
where $p_A^\lambda$ and $p_B^\lambda$ are probabilistic response functions for Alice and Bob respectively. The energy constraint translates from \eqref{Cecon} to
\begin{equation} \label{eq:eacc_w}
\sum_\lambda q(\lambda) p^\lambda_A(a = 0|x)\ \geq\ 1- \omega, \qquad \forall x.
\end{equation}

\section{Entanglement in classical communication}
In our investigation of the role of entanglement in energy-restricted PM scenarios, we begin with the case in which the channel is classical. Since the space of $p_{i}(b|x,y)$ is convex for any $i\in\{\rm QQ,QC,CQ,CC\}$, we can fully characterise it by considering hyperplanes, i.e.~linear functionals of the form
\begin{align}\label{eq:W_gen}
\mathcal{W} = \sum_{b,x,y} c_{bxy} p(b|x,y) \ ,
\end{align}
for some arbitrary real coefficients $c_{bxy}$. We now derive a general bound on any choice of  $\mathcal{W}$ in the EAPM scenario with classical communication. 

\begin{result}\label{thm1}
Consider any $\mathcal{W}$ as defined in \eqref{eq:W_gen}. Any entanglement-assisted classical communication protocol operating at energy no more than $\omega$ satisfies the inequality
\begin{align}\label{eq:W_EACC}
\mathcal{W} \leq \sum_{y}\underset{b}{\max}\left\{\sum_{x}c_{bxy}\right\} + \omega \sum_{xy}\max_b\{c_{bxy}\} . 
\end{align}
\end{result}
\begin{proof}
To prove this, let us define the conditional probability distribution $p(a,b|x,(y,z))=\Tr\left(A_{a|x}\otimes M_{b|y,z}\Psi^{AB}\right)$. By post-selecting on $z=a$ we recover the summand in \eqref{EACC}. Hence, 
\begin{align}\label{step}
\!\!\mathcal{W}_{\rm CQ} \!=\!\!\! \sum_{b,x,y}\! c_{bxy}\bigg[\! p(0,b|x,(y,0))\! +\!\!\sum_{a>0} p(a,b|x,(y,a)) \! \bigg] . 
\end{align}
The first term can be upper-bounded with the Fr\'echet inequality \cite{frechet1935,frechet1951},
\begin{align}
\sum_{b,x,y}c_{bxy} &p(0,b|x,(y,0))\\
&\leq \sum_{b,x,y}c_{bxy} p_B(b|y,0)\leq \sum_{y}\underset{b}{\max}\left\{\sum_{x}c_{bxy}\right\}, \nonumber
\end{align}
where $p_B$ is the marginal of $p$. The right-hand-side corresponds to the first term in \eqref{eq:W_EACC}. Next, consider the second term in \eqref{step}. We have
\begin{align}\nonumber
&\sum_{b,a>0}c_{bxy}p(a,b|x,(y,a)) \\\nonumber
&\leq \max_b\{c_{bxy}\}\sum_{b,a>0}p_A(a,b|x,(y,a))  \\
&=\max_b\{c_{bxy}\}\sum_{a>0}p_A(a|x)\ \leq\ \omega\max_b\{c_{bxy}\} ,
\end{align}
where we used the energy restriction in \eqref{Cecon}. Inserting this back into \eqref{step} we obtain the result in \eqref{eq:W_EACC}.
\end{proof}

The above derivation relies only on basic inequalities that hold for all joint probability distributions. Therefore, the statement of Result~\ref{thm1} can be strengthened in the sense that the inequality \eqref{eq:W_gen} holds not only for correlations obtained under shared entanglement but also for correlations obtained from arbitrary shared no-signalling resources.

\subsection{Probabilistic transmission of classical data}

Since it is fully general, Result~\ref{thm1} will not  provide a tight bound for every choice of hyperplane $\mathcal{W}$ or energy $\omega$. However, there are important quantum communication tasks where it does provide a tight bound, and this will shed light on the role of nonlocal resources in these tasks. A relevant example is the probabilistic transmission of classical data. This task is defined as Alice uniformly sampling $x\in\{0,\ldots,n-1\}$  and Bob aiming to output $b=x$ with a single measurement setting (thus we ignore the input $y$). The average success probability of transmission becomes
\begin{equation}\label{figureofmerit}
\mathcal{W}_{n}=\frac{1}{n}\sum_x p(b=x|x).
\end{equation}
From Result~\ref{thm1} we obtain the bound $\mathcal{W}^{\rm CQ}_n\leq \frac{1}{n}+\omega$. We now show that this bound can be saturated without using entanglement, i.e.~a fully classical (CC) model is sufficient. Consequently, entanglement is not a resource for enhancing the performance of this task.

\begin{result}\label{res2}
Shared entanglement cannot enhance the probabilistic transmission of classical data over an energy-restricted classical channel.
\end{result}
\begin{proof}
We construct an explicit classical model that saturates the bound for entanglement-assisted classical models. To this end, let $\lambda\in\{0,1\}$ be distributed according to $q(0)=\omega$ and $q(1)=1-\omega$. 
\begin{itemize}
	\item If $\lambda=0$, Alice sends $a=x$ and Bob outputs $b=a$. This corresponds to $p_A^0(a|x)=\delta_{a,x}$ and $p_B^0(b|a)=\delta_{a,b}$.
	
	\item  If $\lambda=1$, Alice encodes stochastically if $x=0$, specifically as $p_A^1(0|0)=\nu$ and $p_A^1(1|0)=1-\nu$. For $x=1,2,\ldots,n-1$ she always sends $a=0$, i.e.~$p_A^1(0|x)=1$. Bob outputs $b=a\oplus 1$, corresponding to  $p_B^1(b|a)=\delta_{b,a \oplus 1}$. 
\end{itemize}

By construction, this complies with the energy restriction in \eqref{eq:eacc_w} for $x\neq 0$. For $x=0$, the energy restriction becomes $\omega + \nu (1-\omega) \geq 1-\omega$. Thus, the critical allowed value of $\nu$ is $\nu^*=(1-2\omega)/(1-\omega)$. Furthermore, computing the value of the figure of merit in \eqref{figureofmerit}, we obtain $\mathcal{W}^{\rm CC}=\omega+\frac{1}{n}\left(1-\omega\right)\left(2-\nu\right)$. Inserting $\nu=\nu^*$ returns the result  $\mathcal{W}^{\rm CC}_n=\frac{1}{n}+\omega=\mathcal{W}^{\rm CQ}$.
\end{proof}

\subsection{Random access codes}

That entanglement, or general nonlocal resources, have no capacity to enhance classical communication is not limited only to the probabilistic transmission of classical data. It also holds for other long-studied communication tasks. An example of this is random access codes \cite{ambainis2009, Tavakoli2015}, which are tasks where Alice holds a data set $x\equiv x_1\ldots x_m\in\{0,\ldots,d-1\}^m$ and Bob wants to learn the value of the $y$'th element, i.e.~$b=x_y$. The average success probability is
\begin{equation}\label{RAC}
\mathcal{W}_{m,d}=\frac{1}{m d^m}\sum_{x,y}p(b=x_y|x,y).
\end{equation}
In dimension-restricted settings, it is well-known that the use of entanglement leads to advantages over fully classical models \cite{pawlowski2010, Tav2016}. In contrast, we now show that in energy-restricted communication, entanglement provides no advantage over fully classical models. This applies to the low-energy regime, which is most relevant for practical purposes.

\begin{result}\label{res3}
	For any energy below $\omega\leq d^{-m}$,  shared entanglement cannot enhance any $(m,d)$ random access code over an energy-restricted classical channel.
\end{result}
\begin{proof}
Let $\lambda\in\{\emptyset, S\}$, where $S$ is the set of all strings $s\equiv s_1\ldots s_m\in \{0,\ldots,d-1\}^m$. We assign the probability distribution $q(\emptyset)=1-d^m\omega$ and $q(s)=\omega$ for any $s\in S$. This is a valid distribution whenever $\omega \leq d^{-m}$. The communication strategy is given below, with notation $\bar{0}=00\ldots 0$.
\begin{itemize}
	\item If $\lambda=\emptyset$, Alice sends $a=\bar{0}$ and Bob selects $b$ at random.
	\item If $\lambda=s$, Alice checks whether $x=\lambda$. If positive, she sends $a=s$. If negative, she sends $a=\bar{0}$.  Bob outputs $b=a$.
\end{itemize}
One can now verify that for $x=\bar{0}$ the energy is unit whereas for $x\neq \bar{0}$ it saturates the threshold $1-\omega$. Computing the performance in \eqref{RAC} of this fully classical code yields $\mathcal{W}_{m,d}^{\rm CC}=\frac{1}{d}+\left(1-\frac{1}{d}\right)\omega$. This value is identical to the bound obtained for entanglement-assisted schemes via Result~\ref{thm1} from the relevant coefficients $c_{bxy}=\frac{1}{md^m}\delta_{b,x_y}$.
\end{proof}

\section{Entanglement in quantum communication}
It is well-known that in scenarios without entanglement, substituting energy-restricted classical communication for energy-restricted quantum communication leads to  advantages.  The basic setting in which these advantages come about is the task of probabilistic transmission of a bit, corresponding to the figure of merit $\mathcal{W}_2$ defined in \eqref{figureofmerit}. This task is the basis of nearly all works on the topic so far. It was proven in Ref~\cite{VanHimbeeck2017} that  optimal  quantum correlations without using entanglement are given by 
\begin{align}\label{QCcorr}
\mathcal{W}_2^{\rm QC}=\frac{1}{2}\left(\sqrt{1-\omega}+\sqrt{\omega}\right)^2,
\end{align}
for $\omega\leq \frac{1}{2}$. Above this threshold we trivially have $\mathcal{W}_2=1$. 

In this section, we re-examine the probabilistic transmission of a bit when entanglement is introduced between Alice and Bob. In this EAPM scenario, the correlations are given by \eqref{eq:EAPM_pbxy}. The figure of merit in \eqref{figureofmerit} takes the form
\begin{equation}\label{QQ}
\mathcal{W}_2^{\rm QQ}=\frac{1}{2}\sum_{x=0,1}\Tr\left(\left(\Omega^{A\rightarrow C}_x\otimes\openone^B\right)\left[\Psi^{CB}\right]M^{CB}_{x}\right),
\end{equation}
for some shared entangled state $\Psi$, encoding channels $\Omega_x$ and measurement $\{M_x\}_x$ compatible with the energy restriction in \eqref{energycon}. The central question is whether such a model can be used to outperform the correlations in \eqref{QCcorr}, and if so, how to identify the best protocol.

\subsection{No advantage from unitary encoding}
In search of an entanglement-based quantum advantage in the EAPM scenario, a natural first approach is to adopt the type of protocol made standard since the discovery of dense coding. Specifically, Alice receives her half of a pure $\Psi=\ketbra{\Psi}{\Psi}$ and uses a unitary, $U_x$, to encode her information before  relaying her half to Bob. With this type of protocol, the model in \eqref{QQ} simplifies to
\begin{equation}\label{QQunitary}
\mathcal{W}_2^{\rm QQ}=\frac{1}{2}\sum_{x=0,1}\Tr\left(\left(U_x\otimes \openone\right)\ketbra{\Psi}{\Psi}\left(U_x^\dagger \otimes \openone\right)M^{CB}_{x}\right),
\end{equation}
However, as our next result shows, no entanglement-assisted model of this type can outperform the entanglement-unassisted limit in \eqref{QCcorr}.

\begin{result}\label{res4}
No protocol that uses unitary encoding operations can provide an entanglement-based advantage in the probabilistic transmission of a bit.
\end{result}

\begin{proof}
Let Alice perform the unitary encoding $U_x^{A\rightarrow C}$ to her share of the global bipartite state $\ket*{\Psi^{AB}}$. The global state that arrives to Bob after Alice's encoding is $\ket*{\tau_x^{CB}}=(U_x^{A\rightarrow C}\otimes \mathds{1})\ket*{\Psi^{AB}}$. In the task of probabilistic bit transmission, Bob's goal is to maximise the probability of successfully distinguishing the two states $\ket*{\tau_0^{CB}}$ and $\ket*{\tau_1^{CB}}$. This corresponds to  the Helstrom bound \cite{helstrom1969}
\begin{align}\label{eq:helstrom_2}
\mathcal{W}_2=\frac{1}{2}\left(1+\sqrt{1-\abs{\braket{\tau_0^{CB}}{\tau_1^{CB}}}^2}\right) \ .
\end{align}
Thus, to maximise $\mathcal{W}_2$ we must  minimise the overlap
\begin{align}
\abs{\braket{\tau_0^{CB}}{\tau_1^{CB}}}=\abs{\bra{\tau_0^{CB}}\left(V\otimes \mathds{1}\right)\ket{\tau_0^{CB}}}=\abs{\Tr\left(V\tau_0^{C}\right)} \ ,
\end{align}
where $V=U_1 U_0^\dagger$. Let us write $\tau_0^C=\sum_i \lambda_i \ketbra*{\lambda_i}{\lambda_i}$ in its spectral decomposition, and the unitary $V$ in the same eigenbasis, $V=\sum_{i,j}v_{i,j}\ketbra*{\lambda_i}{\lambda_j}$, for some complex coefficients satisfying $\abs*{v_{i,j}}\leq 1$. Then, $\abs*{\braket*{\tau_0^{CB}}{\tau_1^{CB}}}=\abs{\sum_i \lambda_i v_{i,i}}$. Let us now define $R=\lambda_{i^\ast}v_{i^\ast,i^\ast}$ and $Z=\sum_{i\neq i^\ast}\lambda_i v_{i,i}$, where $i^*=\text{argmax}_i \{\lambda_i \}$. For any pair of complex numbers $R$ and $Z$, it holds that
\begin{align} \label{eq:rev_triang}
\abs{R+Z}\geq \abs{\abs{R}-\abs{Z}} \geq \abs{\abs{R} - (1-\lambda_{i^\ast})} \ ,
\end{align}  
where in the last inequality we used that $\abs{Z}\leq \sum_{i\neq i^\ast}\abs{\lambda_i v_{i,i}} \leq \sum_{i\neq i^\ast}\lambda_i=1-\lambda_{i^\ast}$. To bound $\abs*{\braket*{\tau_0^{CB}}{\tau_1^{CB}}}$ with \eqref{eq:rev_triang}, we need to find the extremes of $R$. The energetic restriction at $\tau_0^C$ is $\bra{0}\tau_0^C\ket{0}=\sum_i\lambda_i\abs{\braket{\lambda_i}{0}}^2\geq 1-\omega$, which already implies that $\lambda_{i^\ast}\geq 1-\omega$. At the extreme, the inequality is saturated when $\ket{\lambda_{i^\ast}}=\ket{0}$. Let $\tau^\perp=\sum_{i\neq i^\ast}\lambda_i \ketbra{\lambda_i}{\lambda_i}$, with $\Tr\left(\tau^\perp\right)=\omega$. With this configuration, now the energetic restriction at $\tau_1^C$ reads,
\begin{align}\nonumber\label{stepeq}
&\bra{0}V\tau_0^C V^\dagger \ket{0} = \bra{\lambda_{i^\ast}}V\tau_0^C V^\dagger \ket{\lambda_{i^\ast}} \\
&= \bra{\lambda_{i^\ast}}V\left((1-\omega)\ketbra{\lambda_{i^\ast}}{\lambda_{i^\ast}}+\tau^\perp\right) V^\dagger \ket{\lambda_{i^\ast}}  \\
&=(1-\omega)|\bra{\lambda_{i^\ast}}V\ket{\lambda_{i^\ast}}|^2 + \bra{\lambda_{i^\ast}} V\tau^\perp V^\dagger \ket{\lambda_{i^\ast}} \geq 1-\omega  \ . \nonumber
\end{align}
The first term of the left-hand-side of the last row is equal to $\abs*{\bra{\lambda_{i^\ast}}V\ket{\lambda_{i^\ast}}}^2 = \abs*{v_{i^\ast,i^\ast}}^2$. The second term becomes $\bra{\lambda_{i^\ast}} V\tau^\perp V^\dagger \ket{\lambda_{i^\ast}}=\sum_{i\neq i^\ast}\lambda_i \abs*{v_{i^\ast,i}}^2 \leq \max_{i\neq i^\ast}\{\lambda_i\}\sum_{i\neq i^\ast}\abs*{v_{i^\ast,i}}^2$. Use that $\max_{i\neq i^\ast}\{\lambda_i\}\leq \omega$ and that since $V$ is unitary $\sum_i |v_{j,i}|^2=1$ for any $j$. Hence,  $\sum_{i\neq i^\ast}\lambda_i \abs*{v_{i^\ast,i}}^2 \leq \omega \left(1-\abs*{v_{i^\ast,i^\ast}}^{2}\right)$. Applied to the last row of \eqref{stepeq} we obtain $(1-\omega)\abs*{v_{i^\ast,i^\ast}}^{2}+\omega(1-\abs*{v_{i^\ast,i^\ast}}^{2})\geq 1-\omega$ which simplifies to $\abs*{v_{i^\ast,i^\ast}}^{2}\geq 1$. Hence, $\abs*{v_{i^\ast,i^\ast}}^{2}=1$. Then, we know that $1-\omega\leq\abs{R}\leq 1$. Inserting this into \eqref{eq:rev_triang} we obtain $\abs*{\braket*{\tau_0^{CB}}{\tau_1^{CB}}}  \geq \abs{1-2\omega}$.  Substituting this bound in \eqref{eq:helstrom_2} one recovers \eqref{QCcorr}.
\end{proof}

\subsection{Advantage from irreversible encoding}
Result~\ref{res4} showed that unitary encoding operations cannot reveal any hypothetical advantages originating from shared entanglement in probabilistic bit transmission. Toward finding an advantage, this leaves us with the option of considering CPTP maps that cannot be reversed. Such operations may appear less intuitive to use than unitary operations, because when applied locally they decohere entanglement by inducing noise. Nevertheless, it is not unheard of that non-unitary operations are useful in  EAPM scenarios, but the known examples are based on dimension-restricted communication settings  \cite{Pauwels2022, Vieira2023, Guo2025} and non-standard, specialised, tasks in which the advantages anyway are small compared to those already obtained from unitary encodings. In contrast to this, we now show that non-unitary encoding operations are essential for unlocking the power of entanglement in energy-restricted quantum communication in its basic primitive task. We summarise the results of this section here below.

\begin{result}
Non-unitary encoding operations on shared entanglement imply a better energy-restricted probabilistic bit transmission than any quantum protocol that does not exploit entanglement. The advantages are achieved already with two-qubit entanglement, but they can be further amplified by schemes exploiting two-qutrit entanglement.
\end{result}
In what follows, we show this result by first considering a scheme based on two-qubit entanglement and  then a scheme based on two-qutrit entanglement.

\subsubsection{Qubit entanglement}
\begin{figure}
\centering
  \includegraphics[width=1.0\columnwidth]{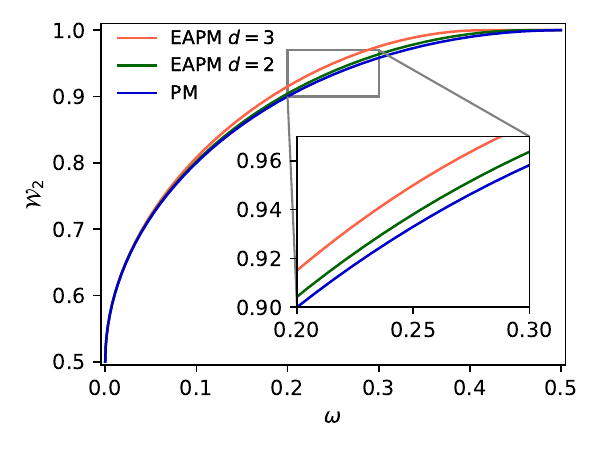}
  \vspace{-0.9cm}
  \caption{\textit{Probabilistic bit transmission.} Achievable value of $\mathcal{W}_2$ at energy at most $\omega$ for quantum messages without entanglement (blue), with two-qubit entanglement (green) and with two-qutrit entanglement (red).}
  \label{fig:SD_bound_2states}
\end{figure}

Let Alice and Bob share a two-qubit state of the form
\begin{align}
\ket{\Psi}_{AB} = \sqrt{1-\omega}\ket{00} - \sqrt{\frac{\omega-r}{2}} \ket{10} + \sqrt{\frac{\omega+r}{2}} \ket{11},
\end{align}
where we have included a free parameter $0\leq r\leq \omega$ that we will specify later.  When $x=0$, Alice's encoding operation is the identity channel. Hence $\tau^{CB}_0=\Psi$. When $x=1$, Alice performs a CPTP map defined by the Kraus operators $K_{0}=\frac{\sqrt{r(\omega-r)}}{\sqrt{(1-\omega)(\omega+r)}}\ketbra{0}{0} + \frac{\sqrt{2r}}{\sqrt{\omega+r}}\ketbra{0}{1}$ and $K_{1}=\frac{\sqrt{1-r-\omega}}{\sqrt{1-\omega}}\ketbra{0}{0}+\frac{\sqrt{2}r}{\sqrt{(1-\omega)(\omega+r)}}\ketbra{1}{0}-\frac{\sqrt{\omega-r}}{\sqrt{\omega+r}}\ketbra{1}{1}$. The state after encoding  becomes $\tau_1^{CB}=\sum_{i=0,1} (K_i\otimes \mathds{1})\Psi (K_i^\dagger\otimes \mathds{1})$ which  can compactly be expressed as $\tau_1^{CB} = r\ketbra{01}{01} + (1-r)\ketbra{\varphi}{\varphi}$ where
\begin{align}
\ket{\varphi}\!=\! \frac{\sqrt{1-r-\omega}}{\sqrt{1-r}}\ket{00}\! +\! \frac{\sqrt{\omega+r}}{\sqrt{2(1-r)}} \ket{10}\! -\! \frac{\sqrt{\omega-r}}{\sqrt{2(1-r)}}\ket{11} .
\end{align}
One can verify that the energy constraint in \eqref{energycon} is saturated for both $\tau^{CB}_0$ and $\tau^{CB}_1$, and that it is independent of $r$. Furthermore, both $\tau^{CB}_0$ and $\tau^{CB}_1$ are entangled since they do not have a positive partial transpose. The optimal value of the figure of merit is given by 
\begin{equation}\label{Helstrom}
\mathcal{W}_2=\frac{1}{2}+ \frac{1}{2}\Tr\left(\left(\tau^{CB}_0-\tau^{CB}_1\right)M_0\right)\leq \frac{1}{2}+ \frac{1}{2}\lambda_+\left(\tau^{CB}_0-\tau^{CB}_1\right),
\end{equation}
where $\lambda_+$ is the sum of the positive eigenvalues. Thus, Bob's optimal measurement is the projection onto the non-negative eigenspace of the operator $\tau_0-\tau_1$. Evaluating this, one finds that there is only one positive eigenvalue,  
\begin{equation}
\frac{1}{2}\! \left(\!r\!+\!\sqrt{\!5 r^2-4 r \omega \!+\!8 \sqrt{z} \sqrt{\omega -r}\! \sqrt{(z-r) (r+\omega )}+8 z \omega }\right),
\end{equation} 
where $z=1-\omega$. This gives a closed analytical form for $\mathcal{W}_2$, which we can now optimise over the free parameter $r$. We have not found a closed form of the solution to that optimisation but it can easily be evaluated for any given $\omega$.  The result is illustrated by the green curve in Fig~\ref{fig:SD_bound_2states}. For every non-trivial $\omega$ the value of $\mathcal{W}_2$ obtained by this strategy is larger than the optimal performance without entanglement given in \eqref{QCcorr}. 

Furthermore, we also optimised $\mathcal{W}_2$ numerically over all possible two-qubit entangled states, CPTP maps for Alice and measurements for Bob. The numerical procedure is an alternating convex search based on semidefinite programming (see the review \cite{SDPrev}). The above analytical construction systematically matches our numerical results, thereby evidencing that it is the optimal protocol based on two-qubit entanglement.

\subsubsection{Qutrit entanglement}
Intuition may suggest that two-qubit entanglement ought to be optimal when Alice has binary inputs. However,  it turns out that entanglement of higher dimension can further amplify the advantage. To that end, consider the following two-qutrit entangled state,
\begin{align}
\ket{\Psi}_{AB} = \sqrt{a}\ket{02} \!+\! \frac{\sqrt{1-a}}{2}\left(\ket{10}-\ket{21}\!+\!\ket{20}\!+\!\ket{11}\right)  ,
\end{align}
with $a=1-r-\omega$, where again $0\leq r\leq \omega$. To this state, Alice applies a CPTP map $\Omega_x^{A\rightarrow C}$ which is represented by Kraus operators $\{K_{x,i}\}_i$. We select these Kraus operators as 
\begin{align}
&K_{x,0} = \sqrt{\frac{r}{1-a}}\left(\ketbra{0}{1}+(-1)^x\ketbra{0}{2}\right) \\
&K_{x,1} = \ketbra{0}{0} - b_x\left(\ketbra{1}{1}+\ketbra{1}{2}\right)+c_x \left(\ketbra{2}{1}-\ketbra{2}{2}\right) \ , \nonumber
\end{align}
where $b_x=(-1)^x\sqrt{\frac{\omega-(-1)^x r}{2\left(1-a\right)}}$ and $c_x=(-1)^x\sqrt{\frac{\omega+(-1)^x r}{2\left(1-a\right)}}$. One can check that these operators satisfy the completeness relation $\sum_i K^\dagger_{x,i}K_{x,i}=\mathds{1}$, $\forall x$. The total state after the encoding can be written as $\tau_x^{CB} = r\ket{0x}\bra{0x} + (1-r)\ket{\psi_x}\bra{\psi_x}$, where
\begin{align}\label{eq:qutrit_states}
\ket{\psi_x} \!=\! \sqrt{\!\frac{a}{1-r}\!}\ket{02} \!-\!\sqrt{\!\frac{1-a}{1-r}\!} b_x \ket{10} \!+\! \sqrt{\!\frac{1-a}{1-r}}c_x\!\ket{21}.
\end{align}
The state of the quantum message takes the simple form $\tau_x^{C}=\left(1-\omega\right)\ketbra{0}{0}+\frac{\omega-(-1)^xr}{2}\ketbra{1}{1}+\frac{\omega+(-1)^xr}{2}\ketbra{2}{2}$, 
from which we see that the energy constraint \eqref{energycon} is saturated. The states $\tau_x^{CB}$ are both entangled and noisy. The former follows from the computable cross norm or realignment (CCNR)
criterion \cite{chen2003,rudolph2005}. The latter is seen from the purity of  $\tau^{CB}_x$, which is $1-2r+r^2$. Hence we induce noise by increasing $r$ in the range $0\leq r\leq \frac{1}{2}$. Moreover, if we consider pure states ($r=0$) the states are partially entangled, as the purity of the reduction $\tau^{C}_x$ is $1-2\omega+3\omega^2/2$. Thus, at low energy, the state is close to a product state.

The optimal value of $\mathcal{W}_2$ is given by inserting our $\tau_x^{CB}$ into \eqref{Helstrom}. The matrix $\tau^{CB}_0-\tau^{CB}_1$ has only four non-zero eigenvalues, and of these only two are positive. One equals $r$ and the other equals $
\sqrt{2a \sqrt{\omega ^2-r^2}+r^2 +2\omega a }$.  Just as in the two-qubit case, this gives us a closed form for $\mathcal{W}_2$ which can easily be optimised over $r$ for any $\omega$.  The result is illustrated by the red curve in \figref{fig:SD_bound_2states}. We see that it improves significantly on the two-qubit model. We remark that also for small $\omega$ there is a strict hierarchy between the red, blue and green curves.

In addition, we have independently employed the alternating convex search method to numerically optimise $\mathcal{W}_2$ over all protocols based on two-qutrit entanglement. The results of this procedure systematically coincide with the analytical result reported above. This indicates that the analytical model is optimal. Moreover, we have also run the same algorithm but with four-dimensional entanglement and found no improvement over the three-dimensional case.

\subsubsection{Correlator space}
\begin{figure}
\centering
  \includegraphics[width=1.0\columnwidth]{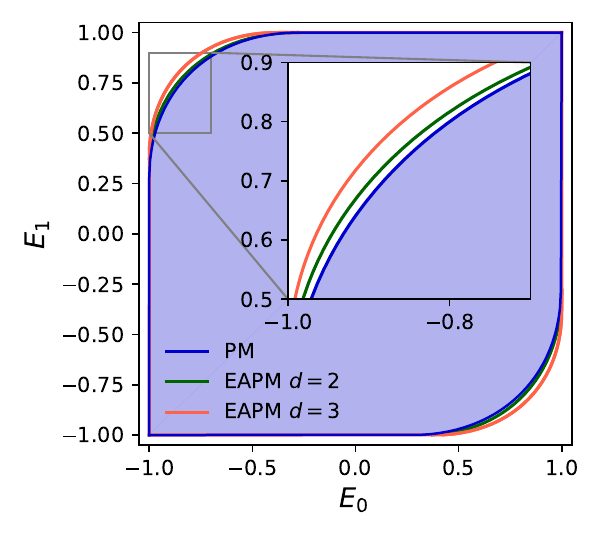}
  \vspace{-0.9cm}
  \caption{\textit{Achievable correlators in quantum models.} The use of entanglement enlarges the space of achievable correlator pairs as compared to the quantum region (blue) reported in \cite{VanHimbeeck2017}. Illustration for $\omega=0.2$.}
  \label{fig:SD_2states}
\end{figure}

Finally, we complete the picture of the achievable correlations for bit transmission with PM and EAPM schemes by considering the correlators $E_x = p(0|x)-p(1|x)$. For a fixed energy bound $\omega$, it has been established that in PM scenarios the set of achievable pairs $(E_0,E_1)$ is the convex hull formed by the points $(1,1)$, $(-1,-1)$ and the ellipse $\left(\frac{E_{+}}{2\gamma}\right)^2 + \left(\frac{E_{-}}{2\sqrt{1-\gamma^2}}\right)^2 = 1$ \cite{VanHimbeeck2017}, where $E_{\pm}=E_0 \pm E_1$ and $\gamma=2\omega-1$. This is illustrated in \figref{fig:SD_2states}. We can complete this picture by considering the pairs achievable with EAPM schemes based on two-qubit and two-qutrit entanglement. We do that through an alternating convex search with semidefinite programs operating in a \textit{seesaw} manner \cite{werner2001}. Specifically, we begin sampling a set of shared states $\tau^{CB}_x$ that fulfil the energy restriction \eqref{energycon} and satisfy $\tau^B_x=\tau^B$. We impose a fixed value for the correlator $E_0$, and optimise the correlator $E_1$ over Bob's measurement. Once a solution is found, we fix the optimal measurement and perform the analogous semidefinite program over the states $\tau^{CB}_x$. This two-step optimisation is iterated until the value of $E_1$ converges. The results are shown in \figref{fig:SD_2states} where we see how entanglement expands the region accessible to energy-constrained quantum models. The result for both the qubit and qutrit case are well-approximated by the optimal discrimination states $\tau_x^{CB}$ reported above.

\begin{figure*}
\centering
  \includegraphics[width=1\textwidth]{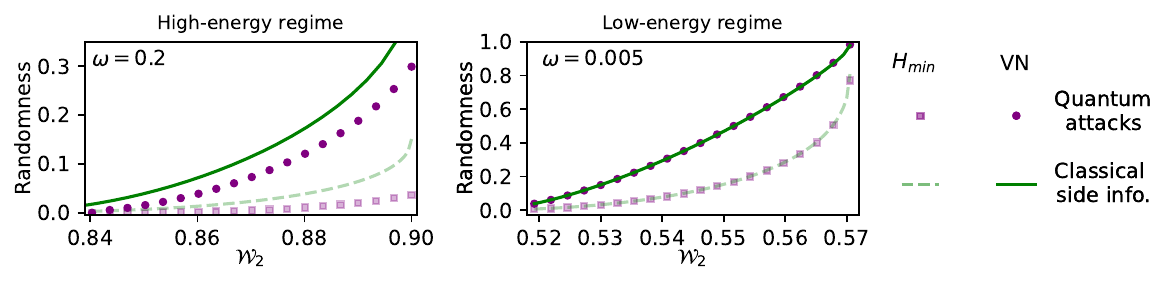}
  \vspace{-0.3cm}
  \caption{\textit{Randomness certification and quantification of quantum side information attacks.} Adversaries with access to pre-shared entanglement with the preparation and measurement devices can harness the correlations that arise in energy-restricted EAPM scenarios to violate lower-bounds on the randomness certified considering classical side information. The impact of these attacks grows in high-energy regimes, but remains negligible at low energies.}
  \label{fig:general}
\end{figure*}

\section{Entanglement in randomness generation}
Probabilistic bit transmission serves as a  primitive for semi-device-independent QRNG operating under energy restrictions. The security of such protocols is established only with respect to classical side information, i.e.~for adversaries that are classically correlated with the devices of Alice and Bob; see e.g.~\cite{vanhimbeeck2019, carceller2025photon}. However, we have shown that the set of quantum correlations is not immune to the presence of entanglement. Consequently, entanglement may play a role in a hypothetical attack on the security of QRNG.  Here, we investigate the role of quantum side information in QRNG and ask whether entanglement can be used by an adversary to undermine the security guarantees of protocols operating under the assumption of classical side information. We investigate this by constructing explicit hacking strategies based on shared entanglement and  we quantify the extent to which the security guarantees are compromised.

\subsection{Randomness certification with classical side information}
When certifying the randomness of measurement outcomes against classical side information, one assumes that a potential adversary may pre-program the devices of both Alice and Bob. The pre-programming is labelled $\lambda$ and it is distributed according to some  probability distribution $q(\lambda)$ known to the adversary but not to Alice and Bob. Thus, in a given round Alice's device prepares a quantum state $\rho_x^{(\lambda)}$, which is sent to Bob who measures ${M_{b|y}^{(\lambda)}}$, obtaining an outcome $b$. Note that $x$ and $y$ are privately selected by Alice and Bob. The probabilities of the variables seen by Alice and Bob are described by \eqref{eq:SRPM_pbxy}. The goal is to show that this distribution  contains randomness also from the point of view of the adversary. The randomness can be quantified by analysing Alice and Bob's observed value of a suitably chosen linear functional $\mathcal{W}$ that now serves as a security parameter.  Assuming that the rounds are independent and identically distributed (i.i.d.), the randomness in Bob's outcome can be quantified through the min-entropy of Bob's output random variable $B$ conditioned on the random variable representing the adversary's side information  $\Lambda_{E}$. It is defined as $H_{\min}(B|\Lambda_{E})=-\log_2(p_g^{\rm Cl})$ \cite{konig2009}, where 
\begin{align}
p_g^{\rm Cl}=\sum_\lambda q(\lambda) \max_b\left\{\Tr\left[\rho_{x^\ast}^{(\lambda)}M_{b|y^{\ast}}^{(\lambda)}\right]\right\},
\end{align} 
which corresponds to the adversary's best probability of guessing Bob's outcome when the inputs are selected as $x=x^\ast$ and $y=y^\ast$. 

In practical scenarios, the i.i.d.~assumption is an unwarranted limitation. To circumvent it, recent works put forward alternative methods to bound the non-i.i.d.~certifiable randomness with the von Neumann entropy \cite{arnon2018,metger2022}. Under classical side information, the von Neumann entropy reduces to the conditional Shannon entropy \cite{carceller2025improving,carceller2025photon}
\begin{align}
\!\!H(B|\Lambda_{E})\! = -\!\!\sum_{\lambda}\!q(\lambda)\!\! \sum_b \!p_{\rm QC}^\lambda(b|x,y) \!\log_2 \!\left(\! p_{\rm QC}^\lambda(b|x,y)\!\right)  .
\end{align}

Both the min- and Shannon entropy of Bob's outcome can be bounded from below using the method of Ref~\cite{carceller2025photon} which is based on semidefinite relaxations of the set of quantum correlations arising under energy restrictions and without shared entanglement. For our probabilistic bit-transmission scenario, we present these two bounds (in green) in \figref{fig:general}: the left-hand plot is for the high-energy regime and the right-hand plot for the low-energy regime. Next, we examine how these bounds hold up to concrete attacks in which the adversary can use quantum side information.

\subsection{Attack with quantum side information}

Consider that the adversary is able to distribute an entangled state to the devices of Alice and Bob. Specifically, the adversary samples $\lambda$ and supplies the state $\Psi_{AB}^{(\lambda)}$ between Alice and Bob. We denote by $\tau_x^\lambda=\left(\Omega_x^{(\lambda)} \otimes\mathds{1}\right)[\Psi^{(\lambda)}] $ the global state after Alice's encoding and by $p_{\rm QQ}^\lambda (b|x)=\Tr\left(\tau_x^{(\lambda)} M_{b}\right)$ the probabilities arising between Alice and Bob conditioned on $\lambda$. Unaware of the entanglement established by the adversary, Alice and Bob design an implementation and observe some value of the security parameter $\mathcal{W}_2^{\rm obs}$ limited by \eqref{QCcorr}. 

In an attempt to increase her guessing probability, the adversary proceeds as follows. With probability $q(0)$ she distributes $\tau_x^0=\{\tau_0^{CB},\tau_1^{CB}\}$; and with probability $q(1)$ she distributes $\tau_x^1=\{\tau_1^{CB},\tau_0^{CB}\}$, where $\tau_x^{CB}$ are the states defined above \eqref{eq:qutrit_states}. Next, she  programs Bob's device to perform the optimal measurement for probabilistic bit transmission, i.e.~$M_0$ is the projector onto the space spanned by the eigenvectors of $(\tau_0^{CB}-\tau_1^{CB})$ associated with positive eigenvalues. This results in the conditional probabilities $p_{\rm QQ}^0(0|0)=p_{\rm QQ}^0(1|1)=\mathcal{W}_{2}^{\rm QQ}$ and $p_{\rm QQ}^1(1|0)=p_{\rm QQ}^1(0|1)=\mathcal{W}_2^{QQ}$. The adversary can then choose the distribution $\{q(0),q(1)\}$  such that the average behaviour matches the success probability seen by Alice and Bob, i.e.~$\sum_\lambda q(\lambda) p_{\rm QQ}^\lambda(x|x)=\mathcal{W}_2^{\rm obs}$ while having a guessing probability of $p_g^{Q}=W_2^{\rm QQ}$.

We generalise the quantum attack for any observable correlation value $\mathcal{W}_2$ maximising the eavesdropper's guessing probability
\begin{align}
p_g^{\rm Q} = \sum_{\lambda}q(\lambda)\max_b\left\{p_{QQ}^\lambda(b|x^\ast)\right\} \ ,
\end{align}
which quantifies the min-entropy conditioned on the adversary's quantum side information $H_{\min}(B|Q_E)=-\log_2(p_g^{\rm Q})$. To compute the quantum attacks, we employ a numerical search algorithm based on semidefinite programs operating in a \textit{seesaw} manner. That is, we iterate the maximisation of $p_g^{\rm Q}$ over the states $\tau_x^{(\lambda)}$ and measurements $M_b$ until a satisfactory convergence is obtained. This yields upper-bounds on the min-entropy of Bob's outcome, which we illustrate in purple square markers in \figref{fig:general}. In the high-energy regime, the computed min-entropies with hacking attacks that harvest shared entanglement lie below the certifiable min-entropy considering classical side information. Thus the attack significantly undermines the security analysis based on classical side information. In contrast, in the low-energy  regime our best found attack  yields only a very small deviation from the classical security analysis.

We observe these qualitative results also for the more relevant case of the conditional von Neumann entropy (see purple circle markers in \figref{fig:general}). To this end, we compute the analogous  attacks based on quantum side information with the von Neumann entropy in EAPM scenarios. As we show in \appref{app.qside_info}, this amounts to 
\begin{align}
H(B|Q_{E}) = -\sum_\lambda q(\lambda) \sum_b p_{\rm QQ}^\lambda(b|x^\ast) \log_2 p_{\rm QQ}^\lambda(b|x^\ast) \ .
\end{align}
We minimise $H(B|Q_{E})$ using an sequential quadratic programming solver implemented via SLSQP from the library Scipy in Python \cite{kraft1988,Virtanen2020}. This algorithm solves a series of quadratic sub-problems to approximate the original nonlinear program. First, we draw valid random initial guesses for states $\tau_x^{(\lambda)}$, measurements $M_b$ and distributions $q(\lambda)$. The cardinality of $\lambda$ is chosen to be the same as the number of measurement outcomes, i.e.~$|\lambda|=2$, although we tried higher values, e.g.~$|\lambda|=3,4$, but saw no difference in the results. Then, at each iteration, the algorithm constructs a local quadratic model of the Lagrangian. The Lagrangian includes the objective $H(B|Q_E)$ and the relevant constraints, namely the  energy restrictions, the  no-signalling in $\tau_x^{(\lambda)}$, the minimum permitted value of $\mathcal{W}_{2}$ and positivity of states $\tau_x^{(\lambda)}\succeq 0$ and measurements $M_{b}\succeq 0$. The resulting quadratic program is solved to determine a search direction that both reduces the objective and satisfies the linearised constraints. The curvature of the Lagrangian is then updated iteratively using quasi-Newton methods, until sufficient convergence is found.

A relevant special case is when the energy bound satisfies $\omega\geq \sqrt{2}-1$. Here, it is not  possible to deterministically transmit the bit in the PM scenario whereas it the EAPM scenario it is possible. As a result, any randomness certified under PM assumptions becomes invalid, since an adversary with quantum side information can fully predict the outcomes. In contrast, at low energies the behaviour of PM and EAPM set-ups becomes nearly indistinguishable, i.e.~$\mathcal{W}^{\rm QC}_{2} \simeq \mathcal{W}^{\rm QQ}_{2}$ and this is reflected in the randomness yield. This is especially relevant because most randomness certification protocols operate in this regime, where quantum states are more indistinguishable and randomness extraction is more efficient. Thus, while our results highlight a potential vulnerability, they also indicate that existing protocols---operating at low energies and assuming only classical side information---remain effectively secure even in the presence of quantum side information.

\section{Discussion}
We have investigated the role of entanglement in energy-restricted communication for the standard prepare-and-measure scenario. Our results show that entanglement plays a significantly different role for these scenarios than it does for the more well-known dimension-restricted communication scenarios.  For classical communication, the latter is known to enhance correlations in a great variety of tasks; a phenomenon that can be linked directly to Bell inequality violations \cite{Buhrman2010}. Here, we have shown that the role of entanglement is much more limited in energy-restricted communication. In particular, we found that two standard classes of tasks, namely probabilistic transmission of classical data and random access codes (for low energy), admit no entanglement-based advantage. It is an interesting open problem whether entanglement is resourceless for every energy-restricted classical communication task, and if not then under what conditions is it a resource?

For quantum communication, we focused on the simplest relevant task and showed that standard unitary encoding operations, commonly used in entanglement-assisted dimension-restricted communication, are resourceless under energy-restrictions. Whether unitary encoding remains resourceless for general tasks is an interesting open problem. Perhaps counter-intuitively, we then showed that non-unitary encoding operations unlock the possibility of an entanglement-based advantage and that the use of systems beyond qubit dimension is relevant for this purpose. In particular, while we found that entanglement improves communication at any non-trivial  energy, we also observed that its impact is significant only beyond the low-energy regime. The low-energy regime is the most relevant for semi-device-independent quantum information experiments since these are often based on weak coherent pulses, which in the low photon number regime accurately approximate a qubit. This motivates the question of whether it could be possible to use simpler experiment set-ups, without entanglement, while still observing quantum correlations strong enough to guarantee nearly undiminished secure yield against adversaries with quantum side information. We investigated this question for quantum random number generation and our best hacking attacks suggests a positive answer. In addition, as our analysis showed that quantum side information can significantly undermine the security analysis when experiments operate outside the low-energy regime. This  can be viewed as a further motivation for focusing implementations in the low-energy regime. Our work leaves the relevant open problem of  establishing cryptographic security proofs for quantum side information: one approach would be to find ways to bound the set of quantum correlations in energy-restricted prepare-and-measure scenarios when sharing arbitrary entanglement.

\begin{acknowledgments}
We thank Jef Pauwels for years of re-occurring discussion about EAPM scenarios with energy constraints.  This work is supported by the Wenner-Gren Foundations, the Knut and Alice Wallenberg Foundation through the Wallenberg Center for Quantum Technology (WACQT) and the Swedish Research Council under Contract No.~2023-03498.
\end{acknowledgments} 

\subsection*{Code availability}
The codes used to generate the results in this paper are available in GitHub: \url{https://github.com/chalswater/Energy_EAPM}.

\subsection*{Note added}
While preparing this manuscript we became aware of a related manuscript by D'Avino \textit{et al.}~which reports similar results. 

\bibliography{quantum_side_info_rng}

\newpage
\onecolumngrid
\appendix

\section{Framework to quantify eavesdropping attacks in quantum randomness certification assuming quantum side information}
\label{app.qside_info}

In this part of the supplemental material we present the framework we consider to quantify the eavesdropping attacks in randomness certification that leverage quantum side information. 

We begin presenting the tripartite scenario including Alice, Bob and Eve. Let all three parties begin sharing a possibly entangled global state $\Psi_{ABE}$. On Alice's side, she privately selects an input $x$ and applies the local channel described by the CPTP map $\Omega_x^{A\rightarrow C}$. Alice sends the output to Bob, who selects an input $y$ and performs a joint measurement described through the POVM $\{M_{b|y}\}_b$ to Alice's output plus his particle, and receives $b$ as an outcome. Afterwards, Eve will measure her particle with the POVM $\{F_{\lambda}\}_{\lambda}$ and receive a result $\lambda$. The overall correlations can be written through the Born rule as
\begin{align} \label{eq:app_corr}
p(b,\lambda|x,y) = \Tr\left[\left(\Omega_{x}^{A\rightarrow C}\otimes \mathds{1}_{BE}\right)\left[\Psi_{ABE}\right]\left(M_{b|y}^{CB}\otimes F_{\lambda}^{E}\right)\right] \ .
\end{align}
As we will see now, the overall correlations can be reinterpreted as follows. Let us split the trace from \eqref{eq:app_corr} into the partial traces $\Tr_{CB}\Tr_{E}$ as,
\begin{align} \label{eq:app_qtauM}
p(b,\lambda|x,y) = \Tr_{CB}\left[\Tr_{E}\left[\left(\Omega_{x}^{A\rightarrow C}\otimes \mathds{1}_{BE}\right)\left[\Psi_{ABE}\right]\left(\mathds{1}_{CB}\otimes F_{\lambda}^{E}\right)\right]M_{b|y}^{CB}\right] = q(\lambda)\Tr\left[\tau_x^{(\lambda)}M_{b|y}\right] \ ,
\end{align}
where we defined 
\begin{align}
q(\lambda) = \Tr\left[\left(\Omega_{x}^{A\rightarrow C}\otimes \mathds{1}_{BE}\right)\left[\Psi_{ABE}\right]\left(\mathds{1}_{CB}\otimes F_{\lambda}^{E}\right)\right] \ \forall x \quad \text{and} \quad \tau_x^{(\lambda)} = \frac{\Tr_{E}\left[\left(\Omega_{x}^{A\rightarrow C}\otimes \mathds{1}_{BE}\right)\left[\Psi_{ABE}\right]\left(\mathds{1}_{CB}\otimes F_{\lambda}^{E}\right)\right]}{\Tr\left[\left(\Omega_{x}^{A\rightarrow C}\otimes \mathds{1}_{BE}\right)\left[\Psi_{ABE}\right]\left(\mathds{1}_{CB}\otimes F_{\lambda}^{E}\right)\right]} \ .
\end{align}
Note that the distribution $q(\lambda)$ does not depend on the input $x$. This follows from non-signalling. Essentially, since Alice's CPTP map acts locally in the subspace $A$ which is not transmitted to Eve, then her measurement outcome $\lambda$ cannot depend on the choice of $x$. The correlations that Alice and Bob observe are then given by $p_{QQ}(b|x,y)=\sum_{\lambda} p(b,\lambda|x,y)$ which are the EAPM correlations specified in the main text. Note that with the re-writing in \eqref{eq:app_qtauM}, the scenario is equivalent to Eve sampling a bi-partite state $\Psi_{AB}^{(\lambda)}=\Tr_{E}\left[\Psi_{ABE}\left(\mathds{1}_{CB}\otimes F_{\lambda}^{E}\right)\right]$, with $\Psi_{ABE}=\sum_\lambda q(\lambda) \Psi_{AB}^{(\lambda)} \otimes \ketbra{\lambda}{\lambda}$, according to $q(\lambda)$ and distributing them to Alice and Bob. Then, after Alice performs the encoding $x$, the bi-partite state is mapped to $\tau_x^{(\lambda)}=\left(\Omega_{x}^{A\rightarrow C}\otimes \mathds{1}_{B}\right)\left[\Psi_{AB}^\lambda\right]$ which Bob will measure with the POVM $\{M_{b|y}\}$. Within this framework, we now present the main figures of merit to quantify the eavesdropping attacks. These are, the min-entropy $H_{\min}(B|Q_{E})$ and the von Neumann entropy $H(B|Q_{E})$. \\

\textbf{Min-entropy.} The min-entropy of Bob's outcome is directly quantified through the eavesdropping guessing probability $p_g^{\rm Q}$, i.e.~$H_{\min}(B|Q_{E})=-\log_2(p_g^{\rm Q})$ \cite{konig2009}. In our framework, $p_g^{\rm Q}$ directly translates to the probability that $\lambda=b$. That is,
\begin{align}
p_g^{\rm Q} = \sum_{\lambda} p(\lambda,\lambda|x^\ast,y^\ast) = \sum_{\lambda} q(\lambda) \Tr\left[\tau_{x^\ast}^{(\lambda)}M_{\lambda|y^\ast}\right] = \sum_{\lambda} q(\lambda) p_{QQ}^\lambda(\lambda|x^\ast,y^\ast) = \sum_{\lambda}q(\lambda)\max_b\left\{p_{QQ}^\lambda(b|x^\ast,y^\ast)\right\} \ ,
\end{align}
where in the last step we reversed-engineered Proposition 1 from Ref.~\cite{bancal2014}. Namely, that the number of strategies $\lambda$ can be reduced to the relevant ones for which the $\max_b$ occurs. These can be labelled as $\lambda=b$, leading to the last equality above.  \\

\textbf{Von Neumann entropy.}  The von Neumann entropy of a system $B$ conditioned on Eve's quantum side information $Q_E$ can be expressed through the unnormalized relative entropy $D(\rho||\sigma)=\Tr\left[\rho\left(\log_2\rho-\log_2\sigma\right)\right]$ as,
\begin{align} \label{eq:app_VN_rel}
H(B|Q_E) = -D\left(\tilde{\Psi}_{x^\ast,y^\ast}||\mathds{1}\otimes \bar{\sigma}_{x^\ast,y^\ast}\right) \ ,
\end{align}
where $\tilde{\Psi}_{x^\ast,y^\ast}$ is the classical-quantum state resulting after Bob's measurement and $\bar{\sigma}_{x^\ast,y^\ast}$ is the averaged state left to Eve after Bob's measurement. Concretely, these can be written as
\begin{align}
\tilde{\Psi}_{x,y} = \sum_b \ketbra{b}{b}\otimes \sigma_{b|x,y} \quad \text{with} \quad \sigma_{b|x,y}=\Tr_{CB}\left[\left(\Omega_{x}^{A\rightarrow C}\otimes \mathds{1}_{BE}\right)\left[\Psi_{ABE}\right]\left(M_{b|y}^{CB}\otimes\mathds{1}_{E}\right)\right] = \sum_{\lambda}q(\lambda)\Tr\left[\tau_x^{(\lambda)}M_{b|y}\right]\ketbra{\lambda}{\lambda} \ ,
\end{align}
and $\bar{\sigma}_{x,y}=\sum_b \sigma_{b|x,y}$. Substituting in \eqref{eq:app_VN_rel}, one ends up with
\begin{align}
H(B|Q_E) &= - \Tr\left[ \sum_b \ketbra{b}{b}\otimes\sigma_{b|x^\ast,y^\ast}\left(\log_2\left(\sum_b \ketbra{b}{b}\otimes\sigma_{b|x^\ast,y^\ast}\right) - \log_2\left(\sum_b \ketbra{b}{b}\otimes \sum_{b'} \sigma_{b'|x^\ast,y^\ast}\right)\right) \right] \\
&= - \sum_b \Tr\left[ \sigma_{b|x^\ast,y^\ast}\left(\log_2\sigma_{b|x^\ast,y^\ast} - \log_2 \sum_{b'} \sigma_{b'|x^\ast,y^\ast}\right) \right] = -\sum_\lambda q(\lambda)\sum_b\Tr\left[\tau_{x^\ast}^{(\lambda)}M_{b|y^\ast}\right] \log_2 \Tr\left[\tau_{x^\ast}^{(\lambda)}M_{b|y^\ast}\right]   \nonumber \\
&= -\sum_\lambda q(\lambda) \sum_b p_{QQ}^\lambda(b|x^\ast,y^\ast) \log_2 p_{QQ}^\lambda(b|x^\ast,y^\ast)  \ , \nonumber
\end{align}
leading to the expression we use in the main text.

\end{document}